\documentstyle[psfig]{caosp}
\begin{document}
%
%
\pubyear{1998}
\volume{23}
\firstpage{1}
\htitle{A0 type stars in the HR diagram}
\hauthor{M. Gerbaldi {\it et al.}}
\title{Determination of $T_{\rm eff}$ and $\log g$ for A0 type stars and its 
impact on the interpretation of the HR diagram with the HIPPARCOS results}
\author{M. Gerbaldi \inst{1,} \inst{2} \and R. Faraggiana
\inst{3}   \and R. Burnage \inst{4} \and F. Delmas \inst{1} \and
A.~Gomez \inst{5} \and S. Grenier \inst{5}}
\institute{Institut d'Astrophysique - 98bis, Bd. Arago - 75014 Paris - France
\and Universit\'e de Paris-Sud XI \and Dipartimento di Astronomia - v. 
Tiepolo 11 - 34131 Trieste - Italy \and Observatoire de Haute-Provence - 
04870 Saint Michel l'Observatoire - France \and DASGAL, 
Observatoire de Meudon - 92150 Meudon  - France}
\date{\today}
\maketitle
\begin{abstract}
The determination of the ``Fundamental Parameters''  $T_{\rm eff}$ and  $\log g$
for a set of dwarf A0-type stars is discussed in terms of consistency
when comparing these values determined through different methods. 
The position of these stars in the HR diagram are discussed, taking into
account the HIPPARCOS data. A large number of binary stars with components 
of similar spectral types has been found from this spectroscopic survey. 
\end{abstract}

\section{Introduction}

\label{intr}
Our purpose is to analyse a set of A0-type, non-giant stars, in order to
define a
sample to be used as {\it standard stars} for further studies. These stars have
parallaxes mesured from the HIPPARCOS experiment and their position in the HR
diagram will be used to derive their evolutionary status.
The spectra obtained in the H$_{\gamma}$ region are used to detect peculiarities
as well as to check the fit with theoretical spectra computed with 
$T_{\rm eff}$ and $\log g$ values determined from calibrations
of colour indices.

\section{Selection of the sample}

All stars with a spectral type A0V in the Bright Stars Catalogue (1982 ed.)
(hereafter BSC),
excluding those recognized as peculiar, shell stars and SB2 have been
selected; this sample contains 230 objects.
Spectroscopic data refer to a second sample of A0 stars brighter than 
the 7th magnitude extracted from a larger set of A-type stars defined in the
framework of our ESO Key programme (Gerbaldi et al. 1989).
Slightly more than 70 A0 type stars were extracted from this set, 
for which spectra were obtained in the spectral
range 4200 - 4500 \AA\  using the Echelec spectrograph attached to the 1.52-m
ESO telescope. All the observations were done at the European Southern 
Observatory (ESO), La Silla, Chile.
The already known peculiar stars, Ap and Am, were a priori excluded from 
the sample.

\vspace{-2mm}
\section{Atmospheric Parameters}

The atmospheric parameters $T_{\rm eff}$ and  $\log g$ were derived from
calibrated colour indices because we had not enough data to derive
$T_{\rm eff}$ from the Infrared flux method (Blackwell et al. 1980).
Very few attention has been given up to now to the dwarf stars in the range
of spectral types A0 -- A3 where the Balmer line profiles vary with both 
$T_{\rm eff}$ and $\log g$.

Any calibration of colour indices in terms of $T_{\rm eff}$ and $\log g$
rely upon :\\
${\bullet}$ synthetic indices \\
These are affected by the ingredients used to compute the atmosphere models
and then the synthetic spectra (chemical abundances, convection...)\\
The synthetic indices are used to define the {\it zero point} by comparing their
values to those of some standard stars.\\
${\bullet}$ {\it well determined} values of $T_{\rm eff}$ and $\log g$ \\
The most commonly used $T_{\rm eff}$ values are those deduced by Code et al. 
(1976) using the angular diameters measured by Hanbury-Brown et al.
(1974) and the energy distribution from the UV to the Infrared.
The $\log g$ values are obtained from binary systems. 

Very recently new measurements of the stellar angular diameters have been done
in Australia (see Booth, 1997), but they have not
been incorporated into the re-calibration of any photometric system.
The accuracy of the Infrared flux method to determine $T_{\rm eff}$ has been
extensively discussed by M\'egessier (1994, 1995, 1997a, 1997b).

The best sources of information for $T_{\rm eff}$ and  $\log g$ for our sample 
of early type stars, are the uvby${\beta}$ and Geneva photometric systems. 
Several calibrations of the uvby${\beta}$ system exist; we shall use that by 
Moon \& Dworetsky (1985) (hereafter quoted MD) which has been 
tested and refined since 
more than a decade.

The MD calibration relies upon synthetic colours and indices from 
Relyea and Kurucz
(1978), Philip and Relyea (1979), Schmidt (1979)
and Kurucz (1979). The zero point has been adjusted by means
of stars with {\it well determined} $T_{\rm eff}$ and $\log g$ values.
The photometric calibration for B-type stars has been re-evaluated by Castelli
(1991). Tests on this calibration done by Balona (1994) and by Napiwotzki
et al. (1993) have shown that no important modification of the MD
calibration is needed for early A-type stars. Napiwotzki et al. (1993) 
suggested a small correction to be
done on $\log g$ for $T_{\rm eff}\geq 9000$~K.
Smalley and Dworetsky (1995) re-evaluated the {\it fundamental}
values of $T_{\rm eff}$ as determined by Code et al. (1976) using recent flux 
measurements; no significant differences were found by these authors.
To conclude, up to now, no revision of the MD calibration of the 
uvby${\beta}$ is required for dwarf A0 stars.

The calibration of the Geneva photometry for early-type stars is very recent
(K\"{u}nzli et al. 1997); 
this system is quite independent from the previous one in the sense that it 
does not include any narrow band centered on a Balmer line; moreover, its 
calibration is done using recent Kurucz flux computations and it is based on a
different and larger set of standard stars. 

It is therefore 
interesting to compare the results obtained from these two systems
and to analyse the coherence of the parameters obtained from different 
aspects of the emitted stellar flux.

\vspace{-2mm}
\section{$T_{\rm eff}$ and $\log g$ accuracy}
\subsection{Influence of the colour excess E(b-y) on $T_{\rm eff}$, $\log g$}
Before computing the atmospheric parameters, the observed colour indices must be
de-reddened. 
For the bright stars of our sample, their visual magnitude being less than 7.0
mag, the reddening is expected to be null or negligible.

\begin{figure}[hbtp]
\psfig{figure=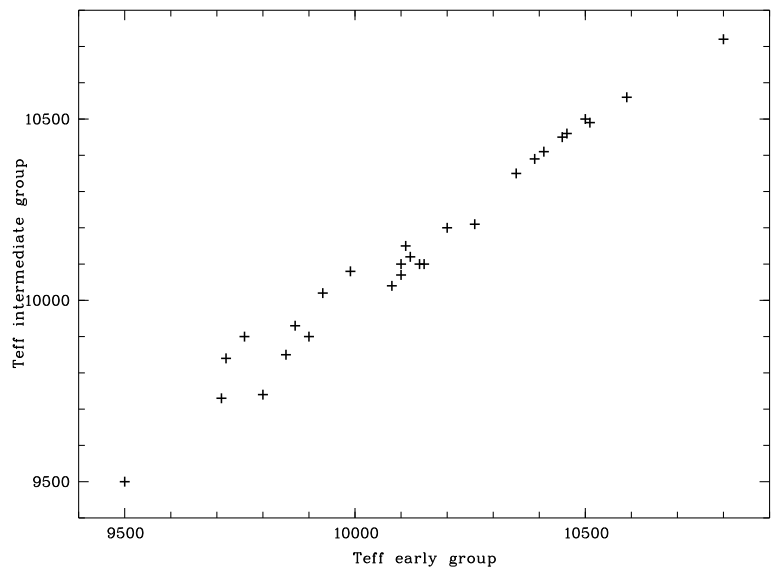,height=8.0cm,angle=-90}
\caption{For the stars with $2.869 \leq \beta \leq 2.880$, 
relation between the $T_{\rm eff}$ computed with the ``early group'' choice and 
the $T_{\rm eff}$ computed with the ``intermediate'' choice}
\label{fp}
\end{figure}

Even though the reddening is expected to be negligible, we tested it;
we used the programme UVBYBETA by Moon (1985) 
in order to check the consistency of the observed values in term of
``normality'' of the star.
We know that any ``blueing'' can be interpreted as due to a Bp-Ap effect, and
any strong reddening requires to have a close look at the spectrum.
The determination of the amount of reddening is based on an empirical
calibration of the uvby${\beta}$ system, so it is free from any atmospheric 
model. Jordi et al. (1997) have discussed the reddening law of Hilditch
(1983) adopted by Moon (1985) ; they have  shown that Moon 
overestimates the E(b-y) value compared to another de-reddening law : the one by
Grosbol (1978). Due to the fact that the stars of our sample have a
small value of their colour excess E(b-y), the error made by using one 
or the other reddening law will be small.

We have shown (Gerbaldi et al. 1997) that a moderate 
reddening of E(b-y)=0.01 produces a difference
in $T_{\rm eff}$ of 200K (see Figure 1 in Ref. cited); on $\log g$ determination,
the effect is negligible (see Figure 2 in Ref. cited).
The stars which are outside the mean relation have values of the parameters
a$_{0}$ and r which are outside the limits of the grids used by the 
MD programme; it is a numerical effect which produces an incorrect result. 

In fact a small non-zero value for E(b-y) can arise from uncertainties on the
colours as well as on the {\it group choice}.
\begin{figure}[hbtp]
\psfig{figure=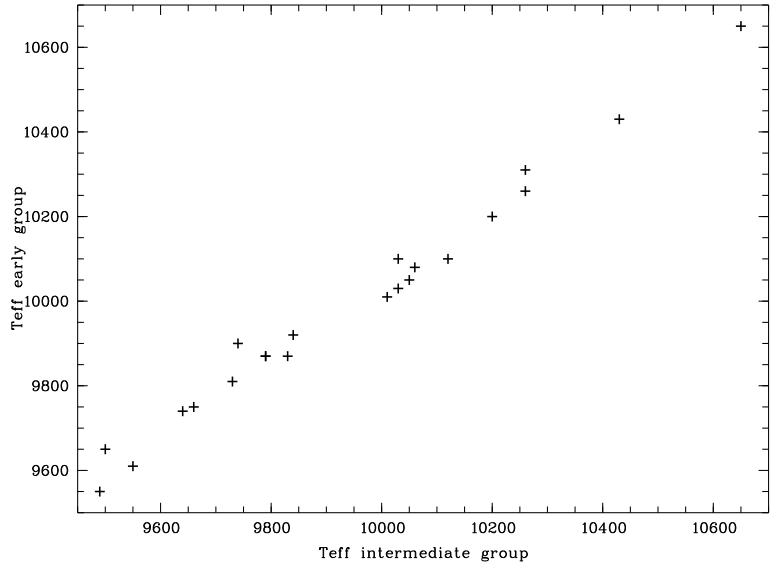,height=8.0cm,angle=-90}
\caption{For the stars 2.880 ${\leq}$ ${\beta}$ ${\leq}$ 2.890, 
relation between the $T_{\rm eff}$ computed with the ``intermediate'' choice
and the $T_{\rm eff}$ computed with the ``early group'' choice}
\label{fp}
\end{figure}

\subsection{Effect of the group choice on $T_{\rm eff}$ and $\log g$}
Dereddening procedures for A0-type stars
require to assign these stars to one of these groups: \\
${\bullet}$ early group : 2.59 ${\leq}$ $\beta$ ${\leq}$ 2.88 \\
${\bullet}$ intermediate group 2.87 ${\leq}$ $\beta$ ${\leq}$ 2.93 \\
In Moon's (1985) programme UVBYBETA the information
concerning the spectral type is added. The assignement
to a group is sometimes ambiguous, errors in group selection cannot
be avoided and this may produce uncertain dereddened colours. 
We have estimated the effect on $T_{\rm eff}$, $\log g$ of a wrong group 
selection, which can be due to an observational error on the value of ${\beta}$.
To select the stars to do such a test, we have considered an error on ${\beta}$
of $\pm 0.01$, so the  ${\beta}$ limits are 2.869 and 2.890.
\begin{figure}[hbtp]
\psfig{figure=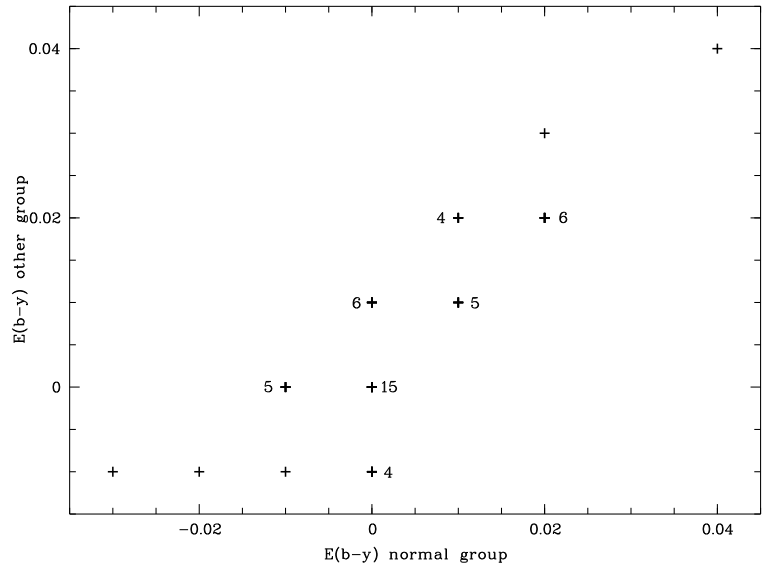,height=8.0cm,angle=-90}
\caption{For the stars with $2.869 \leq \beta \leq 2.880$, 
and $2.880 \leq \beta \leq 2.890$, relation between the color excess
E(b-y) computed in the normal group of the star according to the value 
of $\beta$ and E(b-y) computed in the other group. The number written next to
some points indicates the number of stars having the same values}
\label{fp}
\end{figure}

From our sample of A0V stars we have 28 stars with 
2.869 ${\leq}$ ${\beta}$ ${\leq}$ 2.880 and 22 stars in the range 
2.880 ${\leq}$ ${\beta}$ ${\leq}$ 2.890.
For each selection of stars we computed $T_{\rm eff}$, $\log g$ twice,
that is in each group: early and intermediate.
The results are displayed on Figs. 1 and 2.
For the stars which normally belong to the {\it early} group the computations 
in the {\it intermediate} group give a difference ranging between -70 K
and 80 K.
For the stars which normally belong to the {\it intermediate} group  
the computations  in the {\it early} group give a difference ranging between 
-160 K and 60 K.
The effect on the colour excess E(b-y) is at most of ${\pm} 0.02$ (Fig. 3).

\section{Comparison between $T_{\rm eff}$ and $\log g$ determined from 
uvby${\beta}$ and Geneva Photometry}
Since there is no programme to compute independently the colour excess in the 
Geneva photometric system, we have compared the $T_{\rm eff}$ and $\log g$ 
values only for stars with no reddening. We consider, as above,  normal A0V 
stars from the BSC, but with a further restriction such 
as :  $-0.015 \leq$ E(b-y) $\leq 0.015$.
There are slightly more than 140 objects. For both photometric calibrations the
parameters were computed using E(b-y)=0. 

We note (see Fig. 4) that the values of $T_{\rm eff}$ computed with the MD 
calibration are
systematically higher than those derived from the Geneva photometry calibration
by K\"unzli et al. (1997). The mean value of the difference between these two 
$T_{\rm eff}$ is -70~K with a rms of 137~K.

\begin{figure}[hbtp]
\psfig{figure=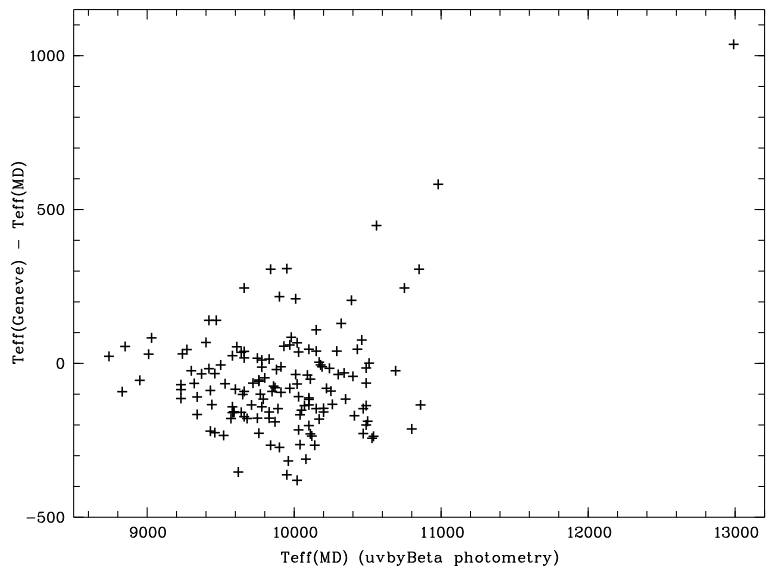,height=8.0cm,angle=-90}
\caption{For the stars with $-0.015 \leq$ E(b-y) $\leq 0.015$, relation
between the $T_{\rm eff}$ computed with the MD calibration and the difference
between the $T_{\rm eff}$ derived from the calibration of the Geneva photometric
system and the $T_{\rm eff}$ from MD}
\label{fp}
\end{figure}
\begin{figure}[hbtp]
\psfig{figure=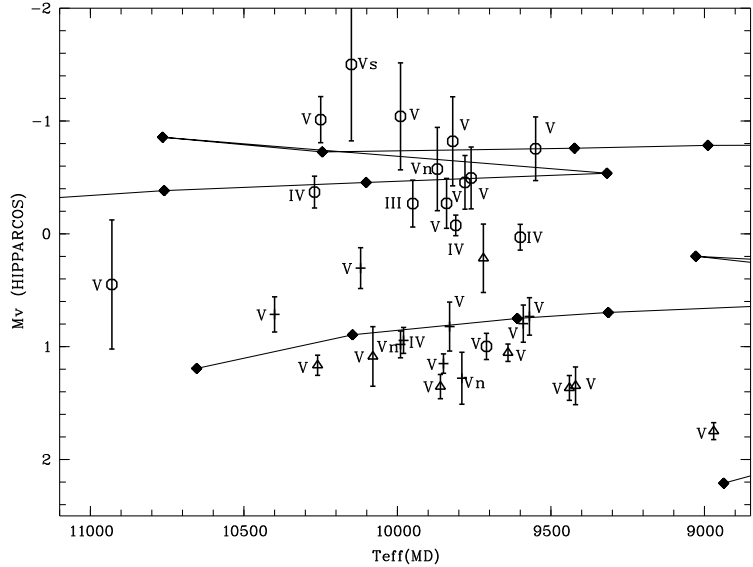,height=8.0cm,angle=-90}
\caption{The HR diagram; the evolutionnary tracks from the top corresponds
respectively to 3.16 M${\odot}$, 2.51 M${\odot}$ and 1.99 M${\odot}$. 
The various symbols correspond to the following values of the gravity : 
hexagon 3.3 ${<}$ $\log g$ ${\leq}$ 3.8; cross : $3.8 <\log g \leq 4.15$;
triangle: 4.15 ${<}$ $\log g$ ${\leq}$ 4.3. The error bars are those given by 
the error on the parallax measurements. The luminosity class is written.}
\label{fp}
\end{figure}
There is also a small systematic effect on the $\log g$ determination: for 
$\log g \leq 3.9$ the value from Geneva photometry is higher than that from
the MD calibration of uvby${\beta}$ colours and it is the reverse for
$\log g > 3.9$.

\vspace{-2mm}
\section{Analysis of the spectroscopic observations}
\vspace{-1mm}
We had spectra at a resolution of 28000 for slightly more than 
70 stars, most of them being classified A0V and the few remaining ones having a 
luminosity class IV or III, or no luminosity class at all.
The reduction of the spectra was done as described by
Burnage \& Gerbaldi (1990, 1992).
Synthetic spectra were computed using the Kurucz solar-abundance models (1993),
and for the atmospheric parameters as computed by the MD programme.
A careful comparison of the synthetic and the observed spectra allows us to
detect a small number of:\\
${\bullet}$ hot Am star \\
${\bullet}$ shell star \\
${\bullet}$ stars with distorded line profiles which are obviously a 
signature of 
binarity.\\
After such a selection we find that 
about half of the stars in the original sample can be considered normal A0 
dwarf stars.

It is striking to note how many binary stars with similar components have been
found from this spectroscopic survey.

\vspace{-2mm}
\section{Conclusion : the HR diagram}
\vspace{-2mm}
With the subsample of normal stars selected above, we have plotted the HR
diagram (Fig. 5). In abscissa, $T_{\rm eff}$ is that from the MD calibration, 
and $M_{\rm V}$ is derived from the HIPPARCOS parallax data. 
We remark that there is no direct relation between the luminosity class 
given by the MK classification and the $M_{\rm V}$ value; stars belonging to 
luminosity class V cover a broad range of $M_{\rm V}$.
The evolutionary tracks by Claret \& Gim\'enez (1992) are overplotted on this 
diagram.


The spread on Fig. 5 points out the difficulty to calibrate
the spectral type A0 and luminosity class V in terms of $M_{\rm V}$.
This spread is so large that it prevents the detection of binaries simply from 
an anomalous position in the HR diagram of stars for which only the spectral
classification is known.

\end{document}